\documentclass[final,1p,times]{elsarticle}
\usepackage{amssymb}
\usepackage{amsfonts}
\usepackage{amsmath}
\usepackage{geometry}
\usepackage{bm}
\numberwithin{equation}{section}

\newdefinition{rmk}{Remark}

\newproof{pf}{Proof}
\newproof{pot}{Proof of Theorem}
\begin{document}

\begin{frontmatter}
\title{The analytic solution for the power series expansion of Heun function}

\author{Yoon Seok Choun\corref{cor1}}
\ead{Yoon.Choun@baruch.cuny.edu; ychoun@gc.cuny.edu; ychoun@gmail.com}
\cortext[cor1]{Correspondence to: Baruch College, The City University of New York, Natural Science Department, A506, 17 Lexington Avenue, New York, NY 10010} 
\address{Baruch College, The City University of New York, Natural Science Department, A506, 17 Lexington Avenue, New York, NY 10010}
\begin{abstract}

The Heun function generalizes all well-known special functions such as Spheroidal Wave, Lame, Mathieu, and hypergeometric $_2F_1$, $_1F_1$ and $_0F_1$ functions. Heun functions are applicable to diverse areas such as theory of black holes, lattice systems in statistical mechanics, solution of the Schr$\ddot{\mbox{o}}$dinger equation of quantum mechanics, and addition of three quantum spins.

In this paper I will apply three term recurrence formula (Choun, Y.S.,  arXiv:1303.0806., 2013) to the power series expansion in closed forms of Heun function (infinite series and polynomial) including all higher terms of $A_n$'s.\footnote{`` higher terms of $A_n$'s'' means at least two terms of $A_n$'s.} Section three contains my analysis on applying the power series expansions of Heun function to a recent paper. (R.S. Maier, Math. Comp. 33, 2007, p.811--843) Due to space restriction final equations for the 192 Heun functions are not included in the paper, but feel free to contact me for the final solutions.  Section four contains two additional examples using the power series expansions of Heun function. 

This paper is 3rd out of 10 in series ``Special functions and three term recurrence formula (3TRF)''. See section 5 for all the papers in the series.  The previous paper in series deals with three term recurrence formula (3TRF). The next paper in the series describes the integral forms of Heun function and its asymptotic behaviors analytically.
\end{abstract}

                                      
\end{frontmatter}  
\section{Introduction}

The recurrence relation without a differential equation dates back to Fibonacci who lived around 1200s. If the birth of modern calculus is attributed to Newton \& Leibniz, then the recurrence relation with a differential equation should also date with the birth of calculus, 350 years ago.

By using the Frobenius method and putting the power series expansion into linear differential equations, the recursive relation of coefficients starts to appear. There can be between two and infinity number of coefficients in the recurrence relation in the power series expansion. During this time, physical phenomena were described using two term recursion relation in the linear ordinary differential equation. More than three terms have been neglected because of its mathematical complexity.

In 1889, K. Heun worked on the second ordinary differential equation which has four regular singular points. The solution was a form of a power series that can be expressed as three term recurrence. Later, the Heun function became a general function of all well-known special functions: Mathieu, Lame and Coulomb spheroidal functions. The coefficients in a power series expansions of Heun equation have a relation between three different coefficients. In contrast, most of well-known special functions consist of two term recursion relation (Hypergeometric, Bessel, Legendre, Kummer functions, etc).   

Due to its complexity the Heun function was neglected for almost 100 years. Recently, the Heun function started to appear in theoretical modern physics, in general relativity, in wave propagation on a background Schwarzschild black hole \cite{rau2004}, in wave equations in curved spaces \cite{Birk2007} and in the Schr$\ddot{\mbox{o}}$dinger equation with anharmonic potential. The Heun function also appears in Kerr-Newman-de Sitter Geometry for massless fields with spin 0 and $\frac{1}{2}$. The angular and radial Teukolsky equations eventually transform into the Heun function \cite{Suzu1998,Suzu1999}
\vspace{3mm}

In this paper I will construct the power series expansion of Heun function in closed forms analytically with the three-term recurrence formula \cite{Chou2012}. Heun's equation is a second-order linear ordinary differential equation of the form \cite{Heun1889,Ronv1995}
\begin{equation}
\frac{d^2{y}}{d{x}^2} + \left(\frac{\gamma }{x} +\frac{\delta }{x-1} + \frac{\epsilon }{x-a}\right) \frac{d{y}}{d{x}} +  \frac{\alpha \beta x-q}{x(x-1)(x-a)} y = 0 \label{eq:1}
\end{equation}
With the condition $\epsilon = \alpha +\beta -\gamma -\delta +1$. The parameters play different roles: $a \ne 0 $ is the singularity parameter, $\alpha $, $\beta $, $\gamma $, $\delta $, $\epsilon $ are exponent parameters, q is the accessory parameter. Also, $\alpha $ and $\beta $ are identical to each other. The total number of free parameters is six. It has four regular singular points which are 0, 1, a and $\infty $ with exponents $\{ 0, 1-\gamma \}$, $\{ 0, 1-\delta \}$, $\{ 0, 1-\epsilon \}$ and $\{ \alpha, \beta \}$. Assume that its solution is
\begin{equation}
y(x)= \sum_{n=0}^{\infty } c_n x^{n+\lambda } \label{eq:2}
\end{equation}
Plug (\ref{eq:2})  into (\ref{eq:1}):
\begin{equation}
c_{n+1}=A_n \;c_n +B_n \;c_{n-1} \hspace{1cm};n\geq 1 \label{eq:3}
\end{equation}
where
\begin{subequations}
\begin{eqnarray}
A_n &=& \frac{(n+\lambda )(n-1+\gamma +\epsilon +\lambda + a(n-1+\gamma +\lambda +\delta ))+q}{a(n+1+\lambda )(n+\gamma +\lambda )}\nonumber\\
&=& \frac{(n+\lambda )(n+\alpha +\beta -\delta +\lambda +a(n+\delta +\gamma -1+\lambda ))+q}{a(n+1+\lambda )(n+\gamma +\lambda )} \label{eq:4a}
\end{eqnarray}
\begin{equation}
B_n = -\frac{(n-1+\lambda )(n+\gamma +\delta +\epsilon -2+\lambda )+\alpha \beta }{a(n+1+\lambda )(n+\gamma +\lambda )}=- \frac{(n-1+\lambda +\alpha )(n-1+\lambda +\beta )}{a(n+1+\lambda )(n+\gamma +\lambda )} \label{eq:4b}
\end{equation}
\begin{equation}
c_1= A_0 \;c_0 \label{eq:4c}
\end{equation}
\end{subequations}
We have two indicial roots which are $\lambda _1= 0$ and $\lambda _2= 1-\gamma $

\section{Power series}
\subsection{Polynomial in which makes $B_n$ term terminated}

In Ref.\cite{Chou2012}, the general expression of power series of $y(x)$ for polynomial of x which makes $B_n$ term terminated is
\begin{eqnarray}
 y(x)&=& \sum_{n=0}^{\infty } y_{n}(x) = y_0(x)+ y_1(x)+ y_2(x)+y_3(x)+\cdots \nonumber\\
&=& c_0 \Bigg\{ \sum_{i_0=0}^{\beta _0} \left( \prod _{i_1=0}^{i_0-1}B_{2i_1+1} \right) x^{2i_0+\lambda } 
+ \sum_{i_0=0}^{\beta _0}\left\{ A_{2i_0} \prod _{i_1=0}^{i_0-1}B_{2i_1+1}  \sum_{i_2=i_0}^{\beta _1} \left( \prod _{i_3=i_0}^{i_2-1}B_{2i_3+2} \right)\right\} x^{2i_2+1+\lambda }\nonumber\\
&&+ \sum_{N=2}^{\infty } \Bigg\{ \sum_{i_0=0}^{\beta _0} \Bigg\{A_{2i_0}\prod _{i_1=0}^{i_0-1} B_{2i_1+1} \prod _{k=1}^{N-1} \Bigg( \sum_{i_{2k}= i_{2(k-1)}}^{\beta _k} A_{2i_{2k}+k}\prod _{i_{2k+1}=i_{2(k-1)}}^{i_{2k}-1}B_{2i_{2k+1}+(k+1)}\Bigg)\nonumber\\
&&\times  \sum_{i_{2N} = i_{2(N-1)}}^{\beta _N} \Bigg( \prod _{i_{2N+1}=i_{2(N-1)}}^{i_{2N}-1} B_{2i_{2N+1}+(N+1)} \Bigg) \Bigg\} \Bigg\} x^{2i_{2N}+N+\lambda }\Bigg\}
  \label{eq:5}
\end{eqnarray}
For a polynomial, we need a condition, which is
\begin{equation}
 B_{2\beta _i + (i+1)}=0 \hspace{1cm} \mathrm{where}\; i= 0,1,2,\cdots, \beta _i=0,1,2,\cdots
 \label{eq:6}
\end{equation}
In this paper the Pochhammer symbol $(x)_n$ is used to represent the rising factorial: $(x)_n = \frac{\Gamma (x+n)}{\Gamma (x)}$.
In the above, $ \beta _i$ is an eigenvalue that makes $B_n$ term terminated at certain value of n. (\ref{eq:6}) makes each $y_i(x)$ where $i=0,1,2,\cdots$ as the polynomial in (\ref{eq:5}).\footnote{In general Heun polynomial comes from a Heun equation that has a fixed integer value of $\alpha $ or $\beta $, just as it has a fixed value of q. In this paper I treat the accessory parameter q as a free variable for the polynomial which makes $B_n$ term terminated.}

\subsubsection{ The case of $\alpha = -2 \alpha _i-i -\lambda $ and $\beta \ne -2 \beta _i -i-\lambda $ where $i, \alpha _i, \beta _i$ = $0,1,2,\cdots$}

In (\ref{eq:4a})-(\ref{eq:4c}) replace $\alpha $ by $-2 \alpha _i-i -\lambda $. In (\ref{eq:6}) replace  index $\beta _i$ by $\alpha _i$. Take the new (\ref{eq:4a})-(\ref{eq:4c}), (\ref{eq:6}) and put them in (\ref{eq:5}).
After the replacement process, the general expression of power series of $y(x)$ for polynomial in which $B_n$ term is terminated is
\begin{eqnarray}
 y(x)&=& \sum_{n=0}^{\infty } y_n(x)= y_0(x)+y_1(x)+y_2(x)+y_3(x)\cdots \nonumber\\
&=& c_0 x^{\lambda } \left\{\sum_{i_0=0}^{\alpha _0} \frac{(-\alpha _0)_{i_0} (\frac{\beta }{2}+\frac{\lambda }{2})_{i_0}}{(1+\frac{\lambda }{2})_{i_0}(\frac{1}{2}+ \frac{\gamma}{2} +\frac{\lambda }{2})_{i_0}} z^{i_0}\right.\nonumber\\
&+& \left\{\sum_{i_0=0}^{\alpha _0}\frac{(i_0+ \frac{\lambda }{2}) \left( i_0+ \Gamma_0^{(S)} \right)+ Q}{(i_0+ \frac{1}{2}+ \frac{\lambda }{2})(i_0 + \frac{\gamma }{2}+ \frac{\lambda }{2})} \frac{(-\alpha _0)_{i_0} (\frac{\beta }{2}+\frac{\lambda }{2})_{i_0}}{(1+\frac{\lambda }{2})_{i_0}(\frac{1}{2}+ \frac{\gamma}{2} +\frac{\lambda }{2})_{i_0}} \right.  \left.\sum_{i_1=i_0}^{\alpha _1} \frac{(-\alpha _1)_{i_1}(\frac{1}{2}+\frac{\beta }{2}+ \frac{\lambda }{2})_{i_1}(\frac{3}{2}+\frac{\lambda }{2})_{i_0}(1+\frac{\gamma }{2}+ \frac{\lambda }{2})_{i_0}}{(-\alpha _1)_{i_0}(\frac{1}{2}+\frac{\beta }{2}+ \frac{\lambda }{2})_{i_0}(\frac{3}{2}+\frac{\lambda }{2})_{i_1}(1+ \frac{\gamma}{2} +\frac{\lambda }{2})_{i_1}} z^{i_1} \right\} \eta \nonumber\\
&+& \sum_{n=2}^{\infty } \left\{ \sum_{i_0=0}^{\alpha _0} \frac{(i_0+ \frac{\lambda }{2}) \left( i_0+ \Gamma_0^{(S)} \right)+ Q}{(i_0+ \frac{1}{2}+ \frac{\lambda }{2})(i_0 + \frac{\gamma }{2}+ \frac{\lambda }{2})}  \frac{(-\alpha _0)_{i_0} (\frac{\beta }{2}+\frac{\lambda }{2})_{i_0}}{(1+\frac{\lambda }{2})_{i_0}(\frac{1}{2}+ \frac{\gamma}{2} +\frac{\lambda }{2})_{i_0}}\right.\nonumber\\
&\times& \prod _{k=1}^{n-1} \left\{ \sum_{i_k=i_{k-1}}^{\alpha _k} \frac{(i_k+\frac{k}{2}+ \frac{\lambda }{2}) \left( i_k+\Gamma_k^{(S)} \right)+ Q}{(i_k+ \frac{k}{2}+\frac{1}{2}+\frac{\lambda }{2})(i_k +\frac{k}{2}+\frac{\gamma }{2}+\frac{\lambda }{2})}  \frac{(-\alpha _k)_{i_k}(\frac{k}{2}+\frac{\beta }{2}+ \frac{\lambda }{2})_{i_k}(1+ \frac{k}{2}+\frac{\lambda }{2})_{i_{k-1}}(\frac{1}{2}+\frac{k}{2}+\frac{\gamma }{2}+ \frac{\lambda }{2})_{i_{k-1}}}{(-\alpha _k)_{i_{k-1}}(\frac{k}{2}+\frac{\beta }{2}+ \frac{\lambda }{2})_{i_{k-1}}(1+\frac{k}{2}+\frac{\lambda }{2})_{i_k}(\frac{1}{2}+ \frac{k}{2}+ \frac{\gamma}{2} +\frac{\lambda }{2})_{i_k}}\right\} \nonumber\\
&\times& \left.\left. \sum_{i_n= i_{n-1}}^{\alpha _n} \frac{(-\alpha _n)_{i_n}(\frac{n}{2}+\frac{\beta }{2}+ \frac{\lambda }{2})_{i_n}(1+ \frac{n}{2}+\frac{\lambda }{2})_{i_{n-1}}(\frac{1}{2}+\frac{n}{2}+\frac{\gamma }{2}+ \frac{\lambda }{2})_{i_{n-1}}}{(-\alpha _n)_{i_{n-1}}(\frac{n}{2}+\frac{\beta }{2}+ \frac{\lambda }{2})_{i_{n-1}}(1+\frac{n}{2}+\frac{\lambda }{2})_{i_n}(\frac{1}{2}+ \frac{n}{2}+ \frac{\gamma}{2} +\frac{\lambda }{2})_{i_n}} z^{i_n} \right\} \eta ^n \right\}\label{eq:7}
\end{eqnarray}
where
\begin{equation}
\begin{cases} z = -\frac{1}{a}x^2 \cr
\eta = \frac{(1+a)}{a} x \cr
\alpha = -2 \alpha _i- i-\lambda \;\;\mbox{as}\;i=0,1,2,\cdots \;\;\mbox{and}\;\; \alpha _i = 0,1,2,\cdots \cr
\alpha _i\leq \alpha _j \;\;\mbox{only}\;\mbox{if}\;i\leq j\;\;\mbox{where}\;i,j= 0,1,2,\cdots
\end{cases}\nonumber  
\end{equation}
and
\begin{equation}
\begin{cases} 
\Gamma_0^{(S)} = \frac{1}{2(1+a)}(-2\alpha _0+ \beta -\delta +a(\delta +\gamma -1+\lambda )) \cr
\Gamma_k^{(S)} = \frac{1}{2(1+a)}(-2\alpha _k+ \beta -\delta +a(\delta +\gamma +\lambda +k-1)) \cr
Q= \frac{q}{4(1+a)}
\end{cases}\nonumber 
\end{equation}
Put $c_0$= 1 as $\lambda $=0 and $\displaystyle{ c_0= a^{-\frac{1}{2}(1-\gamma )}}$ as $\lambda = 1-\gamma $ in (\ref{eq:7}). Then, we obtain two independent solutions of Heun equation. The solution is as follows.
\begin{rmk}
The power series expansion of Heun equation of the first kind for a polynomial in which makes $B_n$ term terminated about $x=0$ as $\alpha = -2 \alpha _j-j $ where $j,\alpha _j \in \mathbb{N}_{0}$ is
\begin{eqnarray}
 y(x)&=& HF_{\alpha _j, \beta }\left( \alpha _j =-\frac{1}{2}(\alpha +j)\big|_{j\in \mathbb{N}_{0}}; \eta = \frac{(1+a)}{a} x ; z= -\frac{1}{a} x^2 \right) \nonumber\\
&=&  \sum_{i_0=0}^{\alpha _0} \frac{(-\alpha _0)_{i_0} (\frac{\beta }{2} )_{i_0}}{(1 )_{i_0}(\frac{1}{2}+ \frac{\gamma}{2} )_{i_0}} z^{i_0} \nonumber\\
&+& \left\{\sum_{i_0=0}^{\alpha _0}\frac{ i_0 \left( i_0+ \Gamma_0^{(S)} \right)+ Q}{(i_0+ \frac{1}{2} )(i_0 + \frac{\gamma }{2})} \frac{(-\alpha _0)_{i_0} (\frac{\beta }{2} )_{i_0}}{(1 )_{i_0}(\frac{1}{2}+ \frac{\gamma}{2} )_{i_0}} \sum_{i_1=i_0}^{\alpha _1} \frac{(-\alpha _1)_{i_1}(\frac{1}{2}+\frac{\beta }{2} )_{i_1}(\frac{3}{2} )_{i_0}(1+\frac{\gamma }{2} )_{i_0}}{(-\alpha _1)_{i_0}(\frac{1}{2}+\frac{\beta }{2} )_{i_0}(\frac{3}{2} )_{i_1}(1+ \frac{\gamma}{2} )_{i_1}} z^{i_1} \right\} \eta \nonumber\\
&+& \sum_{n=2}^{\infty } \left\{ \sum_{i_0=0}^{\alpha _0} \frac{ i_0 \left( i_0+ \Gamma_0^{(S)} \right)+ Q}{(i_0+ \frac{1}{2} )(i_0 + \frac{\gamma }{2} )}  \frac{(-\alpha _0)_{i_0} (\frac{\beta }{2} )_{i_0}}{(1 )_{i_0}(\frac{1}{2}+ \frac{\gamma}{2} )_{i_0}}\right.\nonumber\\
&\times& \prod _{k=1}^{n-1} \left\{ \sum_{i_k=i_{k-1}}^{\alpha _k} \frac{(i_k+\frac{k}{2} ) \left( i_k+\Gamma_k^{(S)} \right)+ Q}{(i_k+ \frac{k}{2}+\frac{1}{2} )(i_k +\frac{k}{2}+\frac{\gamma }{2} )}  \frac{(-\alpha _k)_{i_k}(\frac{k}{2}+\frac{\beta }{2} )_{i_k}(1+ \frac{k}{2} )_{i_{k-1}}(\frac{1}{2}+\frac{k}{2}+\frac{\gamma }{2} )_{i_{k-1}}}{(-\alpha _k)_{i_{k-1}}(\frac{k}{2}+\frac{\beta }{2} )_{i_{k-1}}(1+\frac{k}{2} )_{i_k}(\frac{1}{2}+ \frac{k}{2}+ \frac{\gamma}{2} )_{i_k}}\right\} \nonumber\\
&\times& \left. \sum_{i_n= i_{n-1}}^{\alpha _n} \frac{(-\alpha _n)_{i_n}(\frac{n}{2}+\frac{\beta }{2} )_{i_n}(1+ \frac{n}{2} )_{i_{n-1}}(\frac{1}{2}+\frac{n}{2}+\frac{\gamma }{2} )_{i_{n-1}}}{(-\alpha _n)_{i_{n-1}}(\frac{n}{2}+\frac{\beta }{2} )_{i_{n-1}}(1+\frac{n}{2} )_{i_n}(\frac{1}{2}+ \frac{n}{2}+ \frac{\gamma}{2} )_{i_n}} z^{i_n} \right\} \eta ^n \label{eq:10}
\end{eqnarray}
where
\begin{equation}
\begin{cases} 
\Gamma_0^{(S)} = \frac{1}{2(1+a)}(-2\alpha _0+ \beta -\delta +a(\delta +\gamma -1 )) \cr
\Gamma_k^{(S)} = \frac{1}{2(1+a)}(-2\alpha _k+ \beta -\delta +a(\delta +\gamma +k-1)) \cr
Q= \frac{q}{4(1+a)}
\end{cases}\nonumber 
\end{equation}
\end{rmk}
For the minimum value of Heun equation of the first kind for a polynomial which makes $B_n$ term terminated about $x=0 $, put $\alpha _0=\alpha _1=\alpha _2=\cdots=0$ in (\ref{eq:10}).
\begin{eqnarray}
y(x)&=& HF_{0, \beta }\left( \alpha =-j\big|_{j\in \mathbb{N}_{0}}; \eta = \frac{(1+a)}{a} x ; z= -\frac{1}{a} x^2 \right) \nonumber\\
&=& \sum_{n=0}^{\infty }\frac{\prod_{k=1}^{n}\left( (k-1)\left( k-2+\gamma +\delta +\frac{1}{a}(\beta -\delta )\right)+\frac{q}{a}\right)}{\left( \gamma \right)_n} \frac{x^n}{n!}\nonumber\\
&=& \; _2F_1\left( \frac{ \varphi_1 -\sqrt{ \varphi_1 ^2-\frac{4q}{a}}}{2}, \frac{ \varphi_1 +\sqrt{ \varphi_1 ^2-\frac{4q}{a}}}{2}, \gamma ,x \right) \hspace{1cm}\label{ccc:1}
\end{eqnarray} 
where $\varphi_1 =\gamma +\delta -1+\frac{\beta -\delta }{a}$ and $\left| x\right| < 1$. For the special case, if $x=1$ and $Re\left( -\delta +1+\frac{\delta -\beta }{a}\right)>0 $ in (\ref{ccc:1}),
\begin{eqnarray}
y(x)&=& HF_{0, \beta }\left( \alpha =-j\big|_{j\in \mathbb{N}_{0}}; \eta = \frac{(1+a)}{a} ; z= -\frac{1}{a} \right) \nonumber\\
&=& \frac{ \Gamma \left( \gamma \right) \Gamma \left( \gamma -\varphi_1 \right)}{\Gamma \left( \gamma -\frac{ \varphi_1 -\sqrt{ \varphi_1 ^2-\frac{4q}{a}}}{2}\right) \Gamma \left( \gamma -\frac{ \varphi_1 +\sqrt{ \varphi_1 ^2-\frac{4q}{a}}}{2}\right)}   \nonumber
\end{eqnarray}
\begin{rmk}
The power series expansion of Heun equation of the second kind for a polynomial in which makes $B_n$ term terminated about $x=0$ as $\alpha = -2 \alpha _j-j -1+\gamma  $ where $j,\alpha _j \in \mathbb{N}_{0}$ is
\begin{eqnarray}
y(x)&=& HS_{\alpha _j, \beta }\left( \alpha _j =-\frac{1}{2}(\alpha +1-\gamma +j)\big|_{j\in \mathbb{N}_{0}}; \eta = \frac{(1+a)}{a} x ; z= -\frac{1}{a} x^2 \right) \nonumber\\
&=& z^{\frac{1}{2}(1-\gamma )} \left\{\sum_{i_0=0}^{\alpha _0} \frac{(-\alpha _0)_{i_0} (\frac{\beta }{2}+\frac{1}{2}-\frac{\gamma }{2})_{i_0}}{(\frac{3}{2}-\frac{\gamma }{2})_{i_0}(1)_{i_0}} z^{i_0}\right.\nonumber\\
&+& \left\{\sum_{i_0=0}^{\alpha _0}\frac{(i_0+\frac{1}{2}- \frac{\gamma }{2})\left( i_0+ \Gamma_0^{(S)}\right) +Q}{(i_0+ 1- \frac{\gamma }{2})(i_0 + \frac{1}{2})}\frac{(-\alpha _0)_{i_0} (\frac{\beta }{2}+\frac{1}{2}-\frac{\gamma }{2})_{i_0}}{(\frac{3}{2}- \frac{\gamma }{2})_{i_0}(1)_{i_0}} \right.  \left. \sum_{i_1=i_0}^{\alpha _1} \frac{(-\alpha _1)_{i_1}(\frac{\beta }{2}+1-  \frac{\gamma }{2})_{i_1}(2- \frac{\gamma }{2})_{i_0}(\frac{3}{2})_{i_0}}{(-\alpha _1)_{i_0}(\frac{\beta }{2}+1- \frac{\gamma }{2})_{i_0}(2- \frac{\gamma }{2})_{i_1}(\frac{3}{2})_{i_1}} z^{i_1} \right\} \eta \nonumber\\
&+& \sum_{n=2}^{\infty } \left\{ \sum_{i_0=0}^{\alpha _0} \frac{(i_0+ \frac{1}{2}-\frac{\gamma }{2}) \left( i_0+\Gamma_0^{(S)}\right) +Q}{(i_0+ 1- \frac{\gamma }{2})(i_0 + \frac{1}{2})} \frac{(-\alpha _0)_{i_0} (\frac{\beta }{2}+\frac{1}{2}-\frac{\gamma }{2})_{i_0}}{(\frac{3}{2}-\frac{\gamma }{2})_{i_0}(1)_{i_0}}\right.\nonumber\\
&\times& \prod _{k=1}^{n-1} \left\{ \sum_{i_k=i_{k-1}}^{\alpha _k} \frac{(i_k+\frac{k}{2}+ \frac{1}{2}-\frac{\gamma }{2}) \left( i_k+ \Gamma_k^{(S)}\right) +Q}{(i_k+ \frac{k}{2}+1- \frac{\gamma }{2})(i_k +\frac{k}{2}+\frac{1}{2})}   \frac{(-\alpha _k)_{i_k}(\frac{k}{2}+ \frac{\beta }{2}+\frac{1}{2}-\frac{\gamma }{2})_{i_k}(\frac{k}{2}+\frac{3}{2}-\frac{\gamma }{2})_{i_{k-1}}(\frac{k}{2}+1)_{i_{k-1}}}{(-\alpha _k)_{i_{k-1}}(\frac{k}{2}+ \frac{\beta }{2}+\frac{1}{2}-\frac{\gamma }{2})_{i_{k-1}}(\frac{k}{2}+\frac{3}{2}-\frac{\gamma }{2})_{i_k}(\frac{k}{2}+ 1)_{i_k}}\right\} \nonumber\\
&\times& \left.\left. \sum_{i_n= i_{n-1}}^{\alpha _n} \frac{(-\alpha _n)_{i_n}(\frac{n}{2}+ \frac{\beta }{2}+\frac{1}{2}-\frac{\gamma }{2})_{i_n}(\frac{n}{2}+\frac{3}{2}-\frac{\gamma }{2})_{i_{n-1}}(\frac{n}{2}+1)_{i_{n-1}}}{(-\alpha _n)_{i_{n-1}}(\frac{n}{2}+ \frac{\beta }{2}+\frac{1}{2}-\frac{\gamma }{2})_{i_{n-1}}(\frac{n}{2}+\frac{3}{2}-\frac{\gamma }{2})_{i_n}(\frac{n}{2}+ 1)_{i_n}} z^{i_n} \right\} \eta ^n \right\}\label{eq:11}
\end{eqnarray}
where
\begin{equation}
\begin{cases} 
\Gamma_0^{(S)} = \frac{1}{2(1+a)}(-2\alpha _0+ \beta -\delta +a \delta ) \cr
\Gamma_k^{(S)} = \frac{1}{2(1+a)}(-2\alpha _k+ \beta -\delta +a(\delta +k)) \cr
Q= \frac{q}{4(1+a)}
\end{cases}\nonumber 
\end{equation}
\end{rmk}
For the minimum value of Heun equation of the second kind for a polynomial which makes $B_n$ term terminated about $x=0 $, put $\alpha _0=\alpha _1=\alpha _2=\cdots=0$ in (\ref{eq:11}).
\begin{eqnarray}
y(x)&=& HS_{0, \beta }\left( \alpha =\gamma -1-j\big|_{j\in \mathbb{N}_{0}}; \eta = \frac{(1+a)}{a} x ; z= -\frac{1}{a} x^2 \right) \nonumber\\
&=& \sum_{n=0}^{\infty }\frac{\prod_{k=1}^{n}\left( (k-\gamma )\left( k-1 +\delta +\frac{1}{a}(\beta -\delta )\right)+\frac{q}{a}\right)}{\left( 2-\gamma \right)_n} \frac{x^n}{n!}\nonumber\\
&=& z^{\frac{1}{2}(1-\gamma )}\; _2F_1\left( \frac{\phi_2 -\sqrt{ \varphi_2 ^2-\frac{4q}{a}}}{2}, \frac{\phi_2 +\sqrt{ \varphi_2 ^2-\frac{4q}{a}}}{2}, 2-\gamma ,x \right) \hspace{1cm}\label{ccc:2}
\end{eqnarray} 
where $\phi_2 = \delta -\gamma +1+\frac{1}{a}(\beta -\delta )$, $\varphi_2 =\delta +\gamma -1+\frac{1}{a}(\beta -\delta )$ and $\left| x\right| < 1 $. For the special case, if $x=1$ and $Re\left( -\delta +1+\frac{\delta -\beta }{a}\right)>0 $ in (\ref{ccc:2}),
\begin{eqnarray}
y(x)&=& HS_{0, \beta }\left( \alpha =\gamma -1-j\big|_{j\in \mathbb{N}_{0}}; \eta = \frac{(1+a)}{a} ; z= -\frac{1}{a} \right) \nonumber\\
&=& \frac{ \Gamma \left( 2-\gamma \right) \Gamma \left( 2-\gamma -\phi_2 \right)}{\Gamma \left( 2-\gamma - \frac{\phi_2 -\sqrt{ \varphi_2 ^2-\frac{4q}{a}}}{2}\right) \Gamma \left( 2-\gamma - \frac{\phi_2 +\sqrt{ \varphi_2 ^2-\frac{4q}{a}}}{2}\right)} \left( \frac{-1}{a}\right)^{\frac{1}{2}(1-\gamma )}  \nonumber
\end{eqnarray}
 (\ref{ccc:1}) and (\ref{ccc:2}) tell us that Heun polynomials in which makes $B_n$ term terminated, for fixed value of $\alpha $, require $\left| x\right| < 1$ for the convergence of the radius; it is available for small eigenvalues of $\alpha $. 
(\ref{eq:10}) is the first kind of independent solution of Heun function for the polynomial as $\alpha = -2\alpha_j -j $ where $j=0,1,2,\cdots$. And (\ref{eq:11}) is the second kind of independent solution of Heun function for the polynomial as $\alpha = -2\alpha_j -j -1+\gamma $ where $j=0,1,2,\cdots$. (\ref{eq:7}) is only valid for $\alpha =-2\alpha _i -i- \lambda $ and $\beta \ne -2 \beta _i -i-\lambda $ where $i, \alpha _i, \beta _i = 0,1,2,\cdots$. If $\alpha \ne -2\alpha _i -i- \lambda $ and $\beta = -2 \beta _i -i-\lambda $, replacing $\alpha _i$ and $\beta $ by $\beta _i$ and $\alpha $ in (\ref{eq:7}) because $\alpha $ and $\beta $ are identical to each other.
\subsubsection{The case of $\alpha = -2 \alpha _i-i -\lambda $ and $\beta = -2 \beta _i -i-\lambda $ only if $\alpha _i \leq \beta _i$ where $i, \alpha _i, \beta _i$ = $0,1,2,\cdots$}
 Let $\alpha = -2 \alpha _i-i -\lambda $ and $\beta = -2 \beta _i -i-\lambda $ in (\ref{eq:7}). If $\alpha _i \leq \beta _i$, then index $i_j$ will terminate at the value of $\alpha _i$ where $i,j = 0,1,2,\cdots$ in (\ref{eq:7}). But,
 if $\alpha _i > \beta _i$, then index $i_j$ will terminate at the value of $\beta _i$. Put $\beta = -2\beta _i -i- \lambda $ where $i=0,1,2,\cdots$ in (\ref{eq:7}):
\begin{eqnarray}
 y(x)&=& \sum_{n=0}^{\infty } y_{n}(x)= y_0(x)+y_1(x)+y_2(x)+y_3(x)+\cdots \nonumber\\
&=& c_0 x^{\lambda } \left\{\sum_{i_0=0}^{\alpha _0} \frac{(-\alpha _0)_{i_0} (-\beta _0)_{i_0}}{(1+\frac{\lambda }{2})_{i_0}(\frac{1}{2}+ \frac{\gamma}{2} +\frac{\lambda }{2})_{i_0}} z^{i_0} \right. \nonumber\\
&+& \left\{ \sum_{i_0=0}^{\alpha _0}\frac{(i_0+ \frac{\lambda }{2}) \left( i_0+\Gamma_0^{(B)}\right) + Q}{(i_0+ \frac{1}{2}+ \frac{\lambda }{2})(i_0 + \frac{\gamma }{2}+ \frac{\lambda }{2})}   \frac{(-\alpha _0)_{i_0} (-\beta _0)_{i_0}}{(1+\frac{\lambda }{2})_{i_0}(\frac{1}{2}+ \frac{\gamma}{2} +\frac{\lambda }{2})_{i_0}} \sum_{i_1=i_0}^{\alpha _1} \frac{(-\alpha _1)_{i_1}(-\beta _1)_{i_1}(\frac{3}{2}+\frac{\lambda }{2})_{i_0}(1+\frac{\gamma }{2}+ \frac{\lambda }{2})_{i_0}}{(-\alpha _1)_{i_0}(-\beta _1)_{i_0}(\frac{3}{2}+\frac{\lambda }{2})_{i_1}(1+ \frac{\gamma}{2} +\frac{\lambda }{2})_{i_1}} z^{i_1} \right\} \eta \nonumber\\
&+& \sum_{n=2}^{\infty } \left\{ \sum_{i_0=0}^{\alpha _0} \frac{(i_0+ \frac{\lambda }{2}) \left( i_0+ \Gamma_0^{(B)} \right) + Q}{(i_0+ \frac{1}{2}+ \frac{\lambda }{2})(i_0 + \frac{\gamma }{2}+ \frac{\lambda }{2})}
 \frac{(-\alpha _0)_{i_0} (-\beta _0)_{i_0}}{(1+\frac{\lambda }{2})_{i_0}(\frac{1}{2}+ \frac{\gamma}{2} +\frac{\lambda }{2})_{i_0}}\right. \nonumber\\
&\times& \prod _{k=1}^{n-1} \left\{ \sum_{i_k=i_{k-1}}^{\alpha _k} \frac{(i_k+\frac{k}{2}+ \frac{\lambda }{2}) \left( i_k+\Gamma_k^{(B)}\right) + Q}{(i_k+ \frac{k}{2}+\frac{1}{2}+\frac{\lambda }{2})(i_k +\frac{k}{2}+\frac{\gamma }{2}+\frac{\lambda }{2})}   \frac{(-\alpha _k)_{i_k}(-\beta _k)_{i_k}(1+ \frac{k}{2}+\frac{\lambda }{2})_{i_{k-1}}(\frac{1}{2}+\frac{k}{2}+\frac{\gamma }{2}+ \frac{\lambda }{2})_{i_{k-1}}}{(-\alpha _k)_{i_{k-1}}(-\beta _k)_{i_{k-1}}(1+\frac{k}{2}+\frac{\lambda }{2})_{i_k}(\frac{1}{2}+ \frac{k}{2}+ \frac{\gamma}{2} +\frac{\lambda }{2})_{i_k}}\right\} \nonumber\\
&\times& \left. \left. \sum_{i_n= i_{n-1}}^{\alpha _n} \frac{(-\alpha _n)_{i_n}(-\beta _n)_{i_n}(1+ \frac{n}{2}+\frac{\lambda }{2})_{i_{n-1}}(\frac{1}{2}+\frac{n}{2}+\frac{\gamma }{2}+ \frac{\lambda }{2})_{i_{n-1}}}{(-\alpha _n)_{i_{n-1}}(-\beta _n)_{i_{n-1}}(1+\frac{n}{2}+\frac{\lambda }{2})_{i_n}(\frac{1}{2}+ \frac{n}{2}+ \frac{\gamma}{2} +\frac{\lambda }{2})_{i_n}} z^{i_n} \right\} \eta ^n \right\}\label{eq:13}
\end{eqnarray}
where
\begin{equation}
\begin{cases} z = -\frac{1}{a}x^2 \cr
\eta = \frac{(1+a)}{a} x \cr
\alpha _i\leq \alpha _j \;\;\mbox{only}\;\mbox{if}\;i\leq j\;\;\mbox{where}\;i,j= 0,1,2,\cdots
\end{cases}\nonumber 
\end{equation}
and
\begin{equation}
\begin{cases} 
\Gamma_0^{(B)} = \frac{1}{2(1+a)}(-2\alpha _0-2\beta _0-\delta -\lambda +a(\delta +\gamma -1+\lambda )) \cr
\Gamma_k^{(B)} =  \frac{1}{2(1+a)}(-2\alpha _k-2\beta _k-k-\delta -\lambda  +a(\delta +\gamma +k-1+\lambda )) \cr
Q= \frac{q}{4(1+a)}
\end{cases}\nonumber 
\end{equation}
Put $c_0$= 1 as $\lambda $=0 and $\displaystyle{ c_0= a^{-\frac{1}{2}(1-\gamma )}}$ as $\lambda = 1-\gamma $ in (\ref{eq:13}). Then, we obtain two independent solutions of Heun equation. The solution is as follows.
\begin{rmk}
The power series expansion of Heun equation of the first kind for a polynomial in which makes $B_n$ term terminated about $x=0$ as $\alpha = -2 \alpha _j-j$ and $\beta = -2 \beta _j-j$ where $j,\alpha _j,\beta _j \in \mathbb{N}_{0}$ is
\begin{eqnarray}
y(x)&=& HF_{\alpha _j, \beta_j }\left( \alpha _j =-\frac{1}{2}(\alpha +j), \beta _j =-\frac{1}{2}(\beta +j)\big|_{j \in \mathbb{N}_{0}}; \eta = \frac{(1+a)}{a} x ; z= -\frac{1}{a} x^2 \right) \nonumber\\
&=& \sum_{i_0=0}^{\alpha _0} \frac{(-\alpha _0)_{i_0} (-\beta _0)_{i_0}}{(1 )_{i_0}(\frac{1}{2}+ \frac{\gamma}{2} )_{i_0}} z^{i_0} \nonumber\\
&+& \left\{ \sum_{i_0=0}^{\alpha _0}\frac{ i_0 \left( i_0+\Gamma_0^{(B)}\right) + Q}{(i_0+ \frac{1}{2} )(i_0 + \frac{\gamma }{2} )} \frac{(-\alpha _0)_{i_0} (-\beta _0)_{i_0}}{(1 )_{i_0}(\frac{1}{2}+ \frac{\gamma}{2} )_{i_0}} \sum_{i_1=i_0}^{\alpha _1} \frac{(-\alpha _1)_{i_1}(-\beta _1)_{i_1}(\frac{3}{2} )_{i_0}(1+\frac{\gamma }{2} )_{i_0}}{(-\alpha _1)_{i_0}(-\beta _1)_{i_0}(\frac{3}{2} )_{i_1}(1+ \frac{\gamma}{2} )_{i_1}} z^{i_1} \right\} \eta \nonumber\\
&+& \sum_{n=2}^{\infty } \left\{ \sum_{i_0=0}^{\alpha _0} \frac{ i_0 \left( i_0+ \Gamma_0^{(B)} \right) + Q}{(i_0+\frac{1}{2})(i_0 + \frac{\gamma }{2} )}
 \frac{(-\alpha _0)_{i_0} (-\beta _0)_{i_0}}{(1 )_{i_0}(\frac{1}{2}+ \frac{\gamma}{2} )_{i_0}}\right. \nonumber\\
&\times& \prod _{k=1}^{n-1} \left\{ \sum_{i_k=i_{k-1}}^{\alpha _k} \frac{(i_k+\frac{k}{2} ) \left( i_k+\Gamma_k^{(B)}\right) + Q}{(i_k+ \frac{k}{2}+\frac{1}{2} )(i_k +\frac{k}{2}+\frac{\gamma }{2} )}   \frac{(-\alpha _k)_{i_k}(-\beta _k)_{i_k}(1+ \frac{k}{2} )_{i_{k-1}}(\frac{1}{2}+\frac{k}{2}+\frac{\gamma }{2} )_{i_{k-1}}}{(-\alpha _k)_{i_{k-1}}(-\beta _k)_{i_{k-1}}(1+\frac{k}{2} )_{i_k}(\frac{1}{2}+ \frac{k}{2}+ \frac{\gamma}{2} )_{i_k}}\right\} \nonumber\\
&\times& \left. \sum_{i_n= i_{n-1}}^{\alpha _n} \frac{(-\alpha _n)_{i_n}(-\beta _n)_{i_n}(1+ \frac{n}{2} )_{i_{n-1}}(\frac{1}{2}+\frac{n}{2}+\frac{\gamma }{2} )_{i_{n-1}}}{(-\alpha _n)_{i_{n-1}}(-\beta _n)_{i_{n-1}}(1+\frac{n}{2} )_{i_n}(\frac{1}{2}+ \frac{n}{2}+ \frac{\gamma}{2} )_{i_n}} z^{i_n} \right\} \eta ^n \label{eq:14}
\end{eqnarray}
where
\begin{equation}
\begin{cases} 
\Gamma_0^{(B)} = \frac{1}{2(1+a)}(-2\alpha _0-2\beta _0-\delta +a(\delta +\gamma -1 )) \cr
\Gamma_k^{(B)} =  \frac{1}{2(1+a)}(-2\alpha _k-2\beta _k-k-\delta +a(\delta +\gamma +k-1 )) \cr
Q= \frac{q}{4(1+a)}
\end{cases}\nonumber 
\end{equation}
\end{rmk}
For the minimum value of Heun equation of the first kind for a polynomial which makes $B_n$ term terminated about $x=0 $, put $\alpha _0=\alpha _1=\alpha _2=\cdots=0$ and $\beta _0=\beta _1=\beta _2=\cdots=0$ in (\ref{eq:14}).
\begin{eqnarray}
y(x)&=& HF_{0,0}\left( \alpha =\beta =- j\big|_{j \in \mathbb{N}_{0}}; \eta = \frac{(1+a)}{a} x ; z= -\frac{1}{a} x^2 \right) \nonumber\\
&=& \sum_{n=0}^{\infty }\frac{\prod_{k=1}^{n}\left( (k-1)\left( k-1 + \frac{\delta -a(\gamma +\delta -1)}{(1-a)}\right)-\frac{q}{(1-a)}\right)}{\left( \gamma \right)_n} \frac{\left( \frac{a-1}{a}x\right)^n}{n!}\nonumber\\
&=& \; _2F_1\left( \frac{\varphi _3 -\sqrt{ \varphi _3^2+\frac{4q}{(1-a)}}}{2}, \frac{\varphi _3 +\sqrt{ \varphi _3^2+\frac{4q}{(1-a)}}}{2}, \gamma ,\frac{a-1}{a}x \right) \hspace{1cm}\label{ccc:3}
\end{eqnarray} 
where $\varphi _3 = \frac{1}{(1-a)}(\delta -a(\gamma +\delta -1))$ and $\left|\frac{a-1}{a}x\right| < 1$. For the special case, if $x=\frac{a}{a-1}$ and $Re\left( \gamma -\varphi _3\right)>0 $ in (\ref{ccc:3}),
\begin{eqnarray}
y(x)&=& HF_{0,0}\left( \alpha =\beta =- j\big|_{j \in \mathbb{N}_{0}}; \eta = -1 ; z= -\frac{a}{(1-a)^2} \right) \nonumber\\
&=& \frac{ \Gamma \left( \gamma \right) \Gamma \left( \gamma -\varphi _3  \right)}{\Gamma \left( \gamma - \frac{\varphi _3 -\sqrt{ \varphi _3^2+\frac{4q}{(1-a)}}}{2}\right) \Gamma \left( \gamma - \frac{\varphi _3 +\sqrt{ \varphi _3^2+\frac{4q}{(1-a)}}}{2}\right)}  \nonumber
\end{eqnarray}
\begin{rmk}
The power series expansion of Heun equation of the second kind for a polynomial in which makes $B_n$ term terminated about $x=0$ as $\alpha = -2 \alpha _j-j-1+\gamma  $ and $\beta = -2 \beta _j-j-1+\gamma  $ where $j,\alpha _j,\beta _j \in \mathbb{N}_{0}$ is
\begin{eqnarray}
y(x)&=& HS_{\alpha _j, \beta_j }\left( \alpha _j =-\frac{1}{2}(\alpha +1-\gamma +j), \beta _j =-\frac{1}{2}(\beta +1-\gamma +j)\big|_{j \in \mathbb{N}_{0}}; \eta = \frac{(1+a)}{a} x ; z= -\frac{1}{a} x^2 \right) \nonumber\\
&=& z^{\frac{1}{2}(1-\gamma )} \left\{\sum_{i_0=0}^{\alpha _0} \frac{(-\alpha _0)_{i_0} (-\beta _0)_{i_0}}{(\frac{3}{2}-\frac{\gamma }{2})_{i_0}(1)_{i_0}} z^{i_0} \right. \nonumber\\
&+& \left\{ \sum_{i_0=0}^{\alpha _0}\frac{(i_0+ \frac{1}{2}-\frac{\gamma }{2}) \left( i_0+\Gamma_0^{(B)}\right) + Q}{(i_0+1-\frac{\gamma }{2})(i_0 + \frac{1}{2})}   \frac{(-\alpha _0)_{i_0} (-\beta _0)_{i_0}}{(\frac{3}{2}-\frac{\gamma }{2})_{i_0}(1)_{i_0}} \sum_{i_1=i_0}^{\alpha _1} \frac{(-\alpha _1)_{i_1}(-\beta _1)_{i_1}(2-\frac{\gamma }{2})_{i_0}( \frac{3}{2} )_{i_0}}{(-\alpha _1)_{i_0}(-\beta _1)_{i_0}(2-\frac{\gamma }{2})_{i_1}( \frac{3}{2} )_{i_1}} z^{i_1} \right\} \eta \nonumber\\
&+& \sum_{n=2}^{\infty } \left\{ \sum_{i_0=0}^{\alpha _0} \frac{(i_0+ \frac{1}{2}-\frac{\gamma }{2}) \left( i_0+ \Gamma_0^{(B)} \right) + Q}{(i_0+ 1- \frac{\gamma }{2})(i_0 + \frac{1}{2} )}
 \frac{(-\alpha _0)_{i_0} (-\beta _0)_{i_0}}{(\frac{3}{2}-\frac{\gamma }{2})_{i_0}(1)_{i_0}}\right. \nonumber\\
&\times& \prod _{k=1}^{n-1} \left\{ \sum_{i_k=i_{k-1}}^{\alpha _k} \frac{(i_k+\frac{k}{2}+ \frac{1}{2}-\frac{\gamma }{2}) \left( i_k+\Gamma_k^{(B)}\right) + Q}{(i_k+ \frac{k}{2}+1-\frac{\gamma }{2})(i_k +\frac{k}{2} +\frac{1}{2})}   \frac{(-\alpha _k)_{i_k}(-\beta _k)_{i_k}( \frac{k}{2}+\frac{3}{2}-\frac{\gamma }{2})_{i_{k-1}}( \frac{k}{2}+1)_{i_{k-1}}}{(-\alpha _k)_{i_{k-1}}(-\beta _k)_{i_{k-1}}(\frac{k}{2}+\frac{3}{2}-\frac{\gamma }{2})_{i_k}( \frac{k}{2}+1)_{i_k}}\right\} \nonumber\\
&\times& \left. \left. \sum_{i_n= i_{n-1}}^{\alpha _n} \frac{(-\alpha _n)_{i_n}(-\beta _n)_{i_n}(\frac{n}{2}+\frac{3}{2}-\frac{\gamma }{2})_{i_{n-1}}( \frac{n}{2}+1 )_{i_{n-1}}}{(-\alpha _n)_{i_{n-1}}(-\beta _n)_{i_{n-1}}(\frac{n}{2}+\frac{3}{2}-\frac{\gamma }{2})_{i_n}( \frac{n}{2}+ 1)_{i_n}} z^{i_n} \right\} \eta ^n \right\}\label{eq:15}
\end{eqnarray}
where
\begin{equation}
\begin{cases} 
\Gamma_0^{(B)} = \frac{1}{2(1+a)}(-2\alpha _0-2\beta _0 -1-\delta +\gamma  +a\delta ) \cr
\Gamma_k^{(B)} =  \frac{1}{2(1+a)}(-2\alpha _k-2\beta _k-k-1-\delta +\gamma   +a(\delta +k )) \cr
Q= \frac{q}{4(1+a)}
\end{cases}\nonumber 
\end{equation}
\end{rmk}
For the minimum value of Heun equation of the second kind for a polynomial which makes $B_n$ term terminated about $x=0 $, put $\alpha _0=\alpha _1=\alpha _2=\cdots=0$ and $\beta _0=\beta _1=\beta _2=\cdots=0$ in (\ref{eq:15}).
\begin{eqnarray}
y(x)&=& HS_{0,0}\left( \alpha =\beta =\gamma -1-j\big|_{j \in \mathbb{N}_{0}}; \eta = \frac{(1+a)}{a} x ; z= -\frac{1}{a} x^2 \right) \nonumber\\
&=& z^{\frac{1}{2}(1-\gamma )}\sum_{n=0}^{\infty }\frac{\prod_{k=1}^{n}\left( (k-\gamma )\left( k-1 + \frac{\delta +1-\gamma -a\delta }{(1-a)}\right)-\frac{q}{(1-a)}\right)}{\left( 2-\gamma \right)_n} \frac{\left( \frac{a-1}{a}x\right)^n}{n!}\nonumber\\
&=& z^{\frac{1}{2}(1-\gamma )} \; _2F_1\left( \frac{\phi _4 -\sqrt{ \varphi _4^2 +\frac{4q}{(1-a)}}}{2}, \frac{\phi _4 +\sqrt{ \varphi _4^2 +\frac{4q}{(1-a)}}}{2}, 2-\gamma ,\frac{a-1}{a}x \right) \hspace{1cm}\label{ccc:4}
\end{eqnarray} 
where $\phi _4 =\frac{(2-a)(1-\gamma )}{1-a}+\delta $, $\varphi _4 =\frac{a(1-\gamma )}{1-a}+\delta $ and $\left|\frac{a-1}{a}x\right| < 1$. For the special case, if $x=\frac{a}{a-1}$ and $Re\left( \gamma -\phi _4\right)>0 $ in (\ref{ccc:4}),
\begin{eqnarray} 
y(x)&=& HS_{0,0}\left( \alpha =\beta =\gamma -1-j\big|_{j \in \mathbb{N}_{0}}; \eta = -1 ; z= -\frac{a}{(1-a)^2} \right) \nonumber\\
&=& \frac{ \Gamma \left( 2-\gamma \right) \Gamma \left( 2-\gamma -\phi _4 \right)}{\Gamma \left( 2-\gamma - \frac{\phi _4 -\sqrt{ \varphi _4^2+\frac{4q}{(1-a)}}}{2}\right) \Gamma \left( 2-\gamma - \frac{\phi _4 +\sqrt{ \varphi _4^2+\frac{4q}{(1-a)}}}{2}\right)}  \left( -\frac{a}{(1-a)^2}\right)^{\frac{1}{2}(1-\gamma )} \nonumber
\end{eqnarray}
(\ref{ccc:3}) and (\ref{ccc:4}) inform us that Heun polynomials in which makes $B_n$ term terminated, for fixed values of $\alpha $ and $\beta $, require $\left|\frac{a-1}{a}x\right| < 1$ for the convergence of the radius; it is available for small eigenvalues of $\alpha $ and $\beta $. 
(\ref{eq:14}) is the first kind of independent solution of Heun function for the polynomial as $\alpha = -2\alpha_j -j $ and $\beta = -2\beta _j -j $ only if $\alpha _j \leq \beta _j$ where $j,\alpha_j, \beta_j =0,1,2,\cdots$. And (\ref{eq:15}) is the second kind of independent solution of Heun function for the polynomial as $\alpha = -2\alpha_j -j -1+\gamma $ and $\beta = -2\beta _j -j -1+\gamma $ only if $\alpha _j \leq \beta _j$ where $j,\alpha_j, \beta_j =0,1,2,\cdots$.
\subsection{Infinite series}
In Ref.\cite{Chou2012}, the general expression of power series of $y(x)$ for infinite series is
\begin{eqnarray}
y(x)  &=& \sum_{n=0}^{\infty } y_{n}(x)= y_0(x)+ y_1(x)+ y_2(x)+ y_3(x)+\cdots \nonumber\\
&=& c_0 \Bigg\{ \sum_{i_0=0}^{\infty } \left( \prod _{i_1=0}^{i_0-1}B_{2i_1+1} \right) x^{2i_0+\lambda } 
+ \sum_{i_0=0}^{\infty }\left\{ A_{2i_0} \prod _{i_1=0}^{i_0-1}B_{2i_1+1}  \sum_{i_2=i_0}^{\infty } \left( \prod _{i_3=i_0}^{i_2-1}B_{2i_3+2} \right)\right\} x^{2i_2+1+\lambda }  \nonumber\\
&&+ \sum_{N=2}^{\infty } \Bigg\{ \sum_{i_0=0}^{\infty } \Bigg\{A_{2i_0}\prod _{i_1=0}^{i_0-1} B_{2i_1+1} 
 \prod _{k=1}^{N-1} \Bigg( \sum_{i_{2k}= i_{2(k-1)}}^{\infty } A_{2i_{2k}+k}\prod _{i_{2k+1}=i_{2(k-1)}}^{i_{2k}-1}B_{2i_{2k+1}+(k+1)}\Bigg)\nonumber\\
&&\times  \sum_{i_{2N} = i_{2(N-1)}}^{\infty } \Bigg( \prod _{i_{2N+1}=i_{2(N-1)}}^{i_{2N}-1} B_{2i_{2N+1}+(N+1)} \Bigg) \Bigg\} \Bigg\} x^{2i_{2N}+N+\lambda }\Bigg\} 
\label{eq:19}
\end{eqnarray}
Substitute (\ref{eq:4a})-(\ref{eq:4c}) into (\ref{eq:19}). 
Then, the general expression of power series of $y(x)$ for infinite series is
\begin{eqnarray}
 y(x)&=& \sum_{n=0}^{\infty } y_n(x)= y_0(x)+ y_1(x)+ y_2(x)+ y_3(x)+\cdots \nonumber\\
&=& c_0 x^{\lambda } \left\{\sum_{i_0=0}^{\infty } \frac{(\frac{\alpha }{2}+\frac{\lambda }{2})_{i_0} (\frac{\beta }{2}+\frac{\lambda }{2})_{i_0}}{(1+\frac{\lambda }{2})_{i_0}(\frac{1}{2}+ \frac{\gamma}{2} +\frac{\lambda }{2})_{i_0}} z^{i_0}
+ \left\{\sum_{i_0=0}^{\infty }\frac{(i_0+ \frac{\lambda }{2}) \left( i_0+ \Gamma_0^{(I)}\right)+ Q}{(i_0+ \frac{1}{2}+ \frac{\lambda }{2})(i_0 + \frac{\gamma }{2}+ \frac{\lambda }{2})}\frac{(\frac{\alpha }{2}+\frac{\lambda }{2})_{i_0} (\frac{\beta }{2}+\frac{\lambda }{2})_{i_0}}{(1+\frac{\lambda }{2})_{i_0}(\frac{1}{2}+ \frac{\gamma}{2} +\frac{\lambda }{2})_{i_0}}\right.\right.\nonumber\\
&\times& \left. \sum_{i_1=i_0}^{\infty } \frac{(\frac{1}{2}+\frac{\alpha }{2}+ \frac{\lambda }{2})_{i_1}(\frac{1}{2}+\frac{\beta }{2}+ \frac{\lambda }{2})_{i_1}(\frac{3}{2}+\frac{\lambda }{2})_{i_0}(1+\frac{\gamma }{2}+ \frac{\lambda }{2})_{i_0}}{(\frac{1}{2}+\frac{\alpha }{2}+ \frac{\lambda }{2})_{i_0}(\frac{1}{2}+\frac{\beta }{2}+ \frac{\lambda }{2})_{i_0}(\frac{3}{2}+\frac{\lambda }{2})_{i_1}(1+ \frac{\gamma}{2} +\frac{\lambda }{2})_{i_1}} z^{i_1} \right\} \eta \nonumber\\
&+& \sum_{n=2}^{\infty } \left\{ \sum_{i_0=0}^{\infty } \frac{(i_0+ \frac{\lambda }{2}) \left( i_0+ \Gamma_0^{(I)}\right)+ Q}{(i_0+ \frac{1}{2}+ \frac{\lambda }{2})(i_0 + \frac{\gamma }{2}+ \frac{\lambda }{2})}
 \frac{(\frac{\alpha }{2}+\frac{\lambda }{2})_{i_0} (\frac{\beta }{2}+\frac{\lambda }{2})_{i_0}}{(1+\frac{\lambda }{2})_{i_0}(\frac{1}{2}+ \frac{\gamma}{2} +\frac{\lambda }{2})_{i_0}}\right.\nonumber\\
&\times& \prod _{k=1}^{n-1} \left\{ \sum_{i_k=i_{k-1}}^{\infty } \frac{(i_k+\frac{k}{2}+ \frac{\lambda }{2}) \left( i_k+ \Gamma_k^{(I)}\right)+ Q}{(i_k+ \frac{k}{2}+\frac{1}{2}+\frac{\lambda }{2})(i_k +\frac{k}{2}+\frac{\gamma }{2}+\frac{\lambda }{2})} \right.  \left.\frac{(\frac{k}{2}+\frac{\alpha }{2}+ \frac{\lambda }{2})_{i_k}(\frac{k}{2}+\frac{\beta }{2}+ \frac{\lambda }{2})_{i_k}(1+ \frac{k}{2}+\frac{\lambda }{2})_{i_{k-1}}(\frac{1}{2}+\frac{k}{2}+\frac{\gamma }{2}+ \frac{\lambda }{2})_{i_{k-1}}}{(\frac{k}{2}+\frac{\alpha }{2}+ \frac{\lambda }{2})_{i_{k-1}}(\frac{k}{2}+\frac{\beta }{2}+ \frac{\lambda }{2})_{i_{k-1}}(1+\frac{k}{2}+\frac{\lambda }{2})_{i_k}(\frac{1}{2}+ \frac{k}{2}+ \frac{\gamma}{2} +\frac{\lambda }{2})_{i_k}}\right\} \nonumber\\
&\times& \left.\left. \sum_{i_n= i_{n-1}}^{\infty } \frac{(\frac{n}{2}+\frac{\alpha }{2}+ \frac{\lambda }{2})_{i_n}(\frac{n}{2}+\frac{\beta }{2}+ \frac{\lambda }{2})_{i_n}(1+ \frac{n}{2}+\frac{\lambda }{2})_{i_{n-1}}(\frac{1}{2}+\frac{n}{2}+\frac{\gamma }{2}+ \frac{\lambda }{2})_{i_{n-1}}}{(\frac{n}{2}+\frac{\alpha }{2}+ \frac{\lambda }{2})_{i_{n-1}}(\frac{n}{2}+\frac{\beta }{2}+ \frac{\lambda }{2})_{i_{n-1}}(1+\frac{n}{2}+\frac{\lambda }{2})_{i_n}(\frac{1}{2}+ \frac{n}{2}+ \frac{\gamma}{2} +\frac{\lambda }{2})_{i_n}} z^{i_n} \right\} \eta ^n \right\}\label{eq:20}
\end{eqnarray}
where
\begin{equation}
\begin{cases} 
\Gamma_0^{(I)} =  \frac{1}{2(1+a)}(\alpha +\beta -\delta +\lambda +a(\delta +\gamma -1+\lambda ))\cr
\Gamma_k^{(I)} =  \frac{1}{2(1+a)}(\alpha +\beta -\delta +k +\lambda +a(\delta +\gamma -1+k +\lambda )) \cr
Q= \frac{q}{4(1+a)}
\end{cases}\nonumber 
\end{equation}
Put $c_0$= 1 as $\lambda $=0 and $\displaystyle{ c_0= a^{-\frac{1}{2}(1-\gamma )}}$ as $\lambda = 1-\gamma $ in (\ref{eq:20}). Then, we obtain two independent solutions of Heun equation. The solution is as follows.
\begin{rmk}
The power series expansion of Heun equation of the first kind for an infinite series about $x=0$ using 3TRF is
\begin{eqnarray}
y(x)&=& HF_{\alpha , \beta }\left( \eta = \frac{(1+a)}{a} x ; z= -\frac{1}{a} x^2 \right) \nonumber\\
&=& \sum_{i_0=0}^{\infty } \frac{(\frac{\alpha }{2} )_{i_0} (\frac{\beta }{2} )_{i_0}}{(1 )_{i_0}(\frac{1}{2}+ \frac{\gamma}{2} )_{i_0}} z^{i_0} \nonumber\\
&+& \left\{\sum_{i_0=0}^{\infty }\frac{ i_0 \left( i_0+ \Gamma_0^{(I)}\right)+ Q}{(i_0+ \frac{1}{2} )(i_0 + \frac{\gamma }{2} )}\frac{(\frac{\alpha }{2} )_{i_0} (\frac{\beta }{2} )_{i_0}}{(1 )_{i_0}(\frac{1}{2}+ \frac{\gamma}{2} )_{i_0}}\right. \left. \sum_{i_1=i_0}^{\infty } \frac{(\frac{1}{2}+\frac{\alpha }{2} )_{i_1}(\frac{1}{2}+\frac{\beta }{2} )_{i_1}(\frac{3}{2} )_{i_0}(1+\frac{\gamma }{2} )_{i_0}}{(\frac{1}{2}+\frac{\alpha }{2} )_{i_0}(\frac{1}{2}+\frac{\beta }{2} )_{i_0}(\frac{3}{2} )_{i_1}(1+ \frac{\gamma}{2} )_{i_1}} z^{i_1} \right\} \eta \nonumber\\
&+& \sum_{n=2}^{\infty } \left\{ \sum_{i_0=0}^{\infty } \frac{ i_0 \left( i_0+ \Gamma_0^{(I)}\right)+ Q}{(i_0+ \frac{1}{2} )(i_0 + \frac{\gamma }{2} )}
 \frac{(\frac{\alpha }{2} )_{i_0} (\frac{\beta }{2} )_{i_0}}{(1 )_{i_0}(\frac{1}{2}+ \frac{\gamma}{2} )_{i_0}}\right.\nonumber\\
&\times& \prod _{k=1}^{n-1} \left\{ \sum_{i_k=i_{k-1}}^{\infty } \frac{(i_k+\frac{k}{2} ) \left( i_k+ \Gamma_k^{(I)}\right)+ Q}{(i_k+ \frac{k}{2}+\frac{1}{2} )(i_k +\frac{k}{2}+\frac{\gamma }{2} )} \right.  \left.\frac{(\frac{k}{2}+\frac{\alpha }{2} )_{i_k}(\frac{k}{2}+\frac{\beta }{2} )_{i_k}(1+ \frac{k}{2} )_{i_{k-1}}(\frac{1}{2}+\frac{k}{2}+\frac{\gamma }{2} )_{i_{k-1}}}{(\frac{k}{2}+\frac{\alpha }{2} )_{i_{k-1}}(\frac{k}{2}+\frac{\beta }{2} )_{i_{k-1}}(1+\frac{k}{2} )_{i_k}(\frac{1}{2}+ \frac{k}{2}+ \frac{\gamma}{2} )_{i_k}}\right\} \nonumber\\
&\times& \left. \sum_{i_n= i_{n-1}}^{\infty } \frac{(\frac{n}{2}+\frac{\alpha }{2} )_{i_n}(\frac{n}{2}+\frac{\beta }{2} )_{i_n}(1+ \frac{n}{2} )_{i_{n-1}}(\frac{1}{2}+\frac{n}{2}+\frac{\gamma }{2} )_{i_{n-1}}}{(\frac{n}{2}+\frac{\alpha }{2} )_{i_{n-1}}(\frac{n}{2}+\frac{\beta }{2} )_{i_{n-1}}(1+\frac{n}{2} )_{i_n}(\frac{1}{2}+ \frac{n}{2}+ \frac{\gamma}{2} )_{i_n}} z^{i_n} \right\} \eta ^n  \label{eq:21}
\end{eqnarray}
where
\begin{equation}
\begin{cases} 
\Gamma_0^{(I)} =  \frac{1}{2(1+a)}(\alpha +\beta -\delta +a(\delta +\gamma -1 ))\cr
\Gamma_k^{(I)} =  \frac{1}{2(1+a)}(\alpha +\beta -\delta +k +a(\delta +\gamma -1+k )) \cr
Q= \frac{q}{4(1+a)}
\end{cases}\nonumber 
\end{equation}
\end{rmk}
\begin{rmk}
The power series expansion of Heun equation of the second kind for an infinite series about $x=0$ using 3TRF is
\begin{eqnarray}
y(x)&=& HS_{\alpha , \beta }\left( \eta = \frac{(1+a)}{a} x ; z= -\frac{1}{a} x^2 \right) \nonumber\\
&=& z^{\frac{1}{2}(1-\gamma )} \left\{\sum_{i_0=0}^{\infty } \frac{(\frac{1}{2}+\frac{\alpha }{2}-\frac{\gamma }{2})_{i_0} (\frac{1}{2}+\frac{\beta }{2}-\frac{\gamma }{2} )_{i_0}}{(\frac{3}{2}-\frac{\gamma }{2})_{i_0}(1)_{i_0}} z^{i_0} \right.
\nonumber\\
&+& \left\{\sum_{i_0=0}^{\infty }\frac{(i_0+ \frac{1}{2}-\frac{\gamma }{2}) \left( i_0+ \Gamma_0^{(I)}\right)+ Q}{(i_0+ 1- \frac{\gamma }{2})(i_0 + \frac{1}{2} )}\frac{(\frac{1}{2}+\frac{\alpha }{2}-\frac{\gamma }{2} )_{i_0} (\frac{1}{2}+\frac{\beta }{2}-\frac{\gamma }{2} )_{i_0}}{(\frac{3}{2}-\frac{\gamma }{2})_{i_0}(1)_{i_0}}\right. \left. \sum_{i_1=i_0}^{\infty } \frac{(1+\frac{\alpha }{2}-\frac{\gamma }{2})_{i_1}(1+\frac{\beta }{2}-\frac{\gamma }{2} )_{i_1}(2-\frac{\gamma }{2})_{i_0}(\frac{3}{2})_{i_0}}{(1+\frac{\alpha }{2}-\frac{\gamma }{2} )_{i_0}(1+\frac{\beta }{2}-\frac{\gamma }{2} )_{i_0}(2-\frac{\gamma }{2})_{i_1}(\frac{3}{2})_{i_1}} z^{i_1} \right\} \eta \nonumber\\
&+& \sum_{n=2}^{\infty } \left\{ \sum_{i_0=0}^{\infty } \frac{(i_0+ \frac{1}{2}-\frac{\gamma }{2}) \left( i_0+ \Gamma_0^{(I)}\right)+ Q}{(i_0+ 1-\frac{\gamma }{2})(i_0 + \frac{1}{2})}
 \frac{(\frac{1}{2}+\frac{\alpha }{2}-\frac{\gamma}{2})_{i_0} (\frac{1}{2}+\frac{\beta }{2}-\frac{\gamma}{2} )_{i_0}}{( \frac{3}{2}-\frac{\gamma}{2})_{i_0}(1)_{i_0}}\right.\nonumber\\
&\times& \prod _{k=1}^{n-1} \left\{ \sum_{i_k=i_{k-1}}^{\infty } \frac{(i_k+\frac{k}{2}+ \frac{1}{2}-\frac{\gamma}{2}) \left( i_k+ \Gamma_k^{(I)}\right)+ Q}{(i_k+ \frac{k}{2}+1-\frac{\gamma }{2})(i_k +\frac{k}{2}+\frac{1}{2})} \right.  \left.\frac{(\frac{k}{2}+ \frac{1}{2}+\frac{\alpha }{2}-\frac{\gamma}{2})_{i_k}(\frac{k}{2}+\frac{1}{2}+\frac{\beta }{2}-\frac{\gamma }{2})_{i_k}( \frac{k}{2}+\frac{3}{2}-\frac{\gamma}{2})_{i_{k-1}}( \frac{k}{2}+1)_{i_{k-1}}}{(\frac{k}{2}+\frac{1}{2}+\frac{\alpha }{2}-\frac{\gamma }{2})_{i_{k-1}}(\frac{k}{2}+\frac{1}{2}+\frac{\beta }{2}-\frac{\gamma }{2})_{i_{k-1}}( \frac{k}{2}+\frac{3}{2}-\frac{\gamma}{2})_{i_k}( \frac{k}{2}+1)_{i_k}}\right\} \nonumber\\
&\times& \left.\left. \sum_{i_n= i_{n-1}}^{\infty } \frac{(\frac{n}{2}+ \frac{1}{2}+\frac{\alpha }{2}-\frac{\gamma}{2})_{i_n}(\frac{n}{2}+\frac{1}{2}+\frac{\beta }{2}-\frac{\gamma}{2})_{i_n}( \frac{n}{2}+\frac{3}{2}-\frac{\gamma }{2})_{i_{n-1}}( \frac{n}{2}+1 )_{i_{n-1}}}{(\frac{n}{2}+\frac{1}{2}+\frac{\alpha }{2}-\frac{\gamma }{2})_{i_{n-1}}(\frac{n}{2}+\frac{1}{2}+\frac{\beta }{2}-\frac{\gamma }{2})_{i_{n-1}}( \frac{n}{2}+\frac{3}{2}-\frac{\gamma }{2})_{i_n}( \frac{n}{2}+1)_{i_n}} z^{i_n} \right\} \eta ^n \right\}\label{eq:22}
\end{eqnarray}
where
\begin{equation}
\begin{cases} 
\Gamma_0^{(I)} =  \frac{1}{2(1+a)}(\alpha +\beta -\gamma -\delta +1+a \delta )\cr
\Gamma_k^{(I)} =  \frac{1}{2(1+a)}(\alpha +\beta -\gamma -\delta +k+1+a(\delta +k)) \cr
Q= \frac{q}{4(1+a)}
\end{cases}\nonumber 
\end{equation}
\end{rmk}
(\ref{eq:21}) is the first kind of independent solution of Heun function for the infinite series. And (\ref{eq:22}) is the second kind of independent solution of Heun function for the infinite series. Also, it is required that $\gamma \ne 0,-1,-2,\cdots$ for the first kind of independent solution of Heun function for all cases. Because if it does not, its solution will be divergent. And it's required that $\gamma \ne 2,3,4, \cdots$ for the second kind of independent solution of Heun function for all cases.
\section{Power series analysis of 192 Heun functions}
1. A machine-generated list of 192 (isomorphic to the Coxeter group of the Coxeter diagram $D_4$) local solutions of the Heun equation was obtained by Robert S. Maier(2007) \cite{Maie2007}. By using the three term recurrence formula \cite{Chou2012}, we obtain analytic power series expansion in closed forms of 192 solutions of Heun function for polynomial and infinite series. For example, one of the 192 local solution of Heun function in Table 2 \cite{Maie2007} is
\begin{equation}
(1-x)^{1-\delta } Hl(a, q - (\delta  - 1)\gamma a; \beta  - \delta  + 1, \alpha  - \delta + 1, \gamma ,2 - \delta ; x) \label{eq:61}
\end{equation} 
Replacing coefficients q, $\alpha$, $\beta$, and $\delta$ by $q - (\delta  - 1)\gamma a $, $\beta  - \delta  + 1 $, $\alpha  - \delta + 1$ and $2 - \delta$ into (\ref{eq:7}), (\ref{eq:13}), (\ref{eq:20}), we obtain power series expansion in closed forms of (\ref{eq:61}). 
 
\section{Additional examples of Heun function in black hole problems}

2. In ``Perturbations of Kerr-de Sitter Black Hole and Heun's Equation''\cite{Suzu1998}, ``Analytic solution of Teukolsky Equation in Kerr-de Sitter and Kerr-Newman-de Sitter Geometries''\cite{Suzu1999}, the authors consider the Teukolsky equations for the Kerr-Newman-de Sitter geometries. The Newman-Penrose formalism results into two equations: the angular Teukolsky equation and the radial Teukolsky equation. These two equations are transformed into Heun equations (see (2.15), (3.18), (4.9), (4.12) in Ref.\cite{Suzu1998} and (3.4), (3.9), (4.40) in Ref.\cite{Suzu1999}). 
Using the power series expansion of Heun equation, it might be possible to obtain specific eigenvalues for each individual region, for the entire region of r .
Using the integral forms of Heun equation, derived in \cite{Chou2012d}, it might be possible to construct the normalized wave functions and the relative scattering probability (absorption rate, decay rate) for each individual region (satisfying some specific boundary conditions), for the entire region of r. 

\vspace{5mm}
3. In ``Quantized black hole and Heun function''\cite{Mome2012}, the assumption made is that the black hole behaves like a particle with mass M, therefore allowing the radial wave equation for the quantized black hole be transformed into Heun function (see (6), (8) in Ref.\cite{Mome2012}). 

Using power series expansions of Heun equation, it might be possible to obtain eigenvalues at various regions. Using integral forms \cite{Chou2012d} of Heun function it might be possible to obtain normalized constants at various regions. Integral forms of Heun function and its analytic boundary condition are important because the results can be extracted into special functions such as the Mathieu, Lame, Spheroidal Wave and hypergeometric $_2F_1$, $_1F_1$ and $_0F_1$ functions and etc. In \cite{Chou2012d}, I am investigating integral forms of Heun function and its asymptotic behaviors analytically. 
\vspace{5mm}

\section{Series ``Special functions and three term recurrence formula (3TRF)''} 

This paper is 3rd out of 10.
\vspace{3mm}

1. ``Approximative solution of the spin free Hamiltonian involving only scalar potential for the $q-\bar{q}$ system'' \cite{Chou2012a}--in order to solve the spin-free Hamiltonian with light quark masses we are led to develop a totally new kind of special function theory in mathematics that generalize all existing theories of confluent hypergeometric types. We call it the Grand Confluent Hypergeometric Function. Our new solution produces previously unknown extra hidden quantum numbers relevant for the description of supersymmetry and for generating new mass formulas.
\vspace{3mm}

2. ``Generalization of the three-term recurrence formula and its applications'' \cite{Chou2012b}--generalize the three term recurrence formula in the linear differential equation. Obtain the exact solution of the three term recurrence for polynomials and infinite series.
\vspace{3mm}

3. ``The analytic solution for the power series expansion of Heun function'' \cite{Chou2012c}--apply the three term recurrence formula to the power series expansion in closed forms of Heun function (infinite series and polynomials) including all higher terms of $A_n$'s.
\vspace{3mm}

4. ``Asymptotic behavior of Heun function and its integral formalism'', \cite{Chou2012d}--apply the three term recurrence formula, derive the integral formalism, and analyze the asymptotic behavior of Heun function (including all higher terms of $A_n$'s). 
\vspace{3mm}

5. ``The power series expansion of Mathieu function and its integral formalism'', \cite{Chou2012e}--apply  the three term recurrence formula, and analyze the power series expansion of Mathieu function and its integral forms.  
\vspace{3mm}

6. ``Lame equation in the algebraic form'' \cite{Chou2012f}--apply the three term recurrence formula, and analyze the power series expansion of Lame function in the algebraic form and its integral forms.
\vspace{3mm}

7. ``Power series and integral forms of Lame equation in Weierstrass's form and its asymptotic behaviors'' \cite{Chou2012g}--apply the three term recurrence formula, and derive the power series expansion of Lame function in Weierstrass's form and its integral forms. 
\vspace{3mm}

8. ``The generating functions of Lame equation in Weierstrass's form'' \cite{Chou2012h}--derive the generating functions of Lame function in Weierstrass's form (including all higher terms of $A_n$'s). Apply integral forms of Lame functions in Weierstrass's form.
\vspace{3mm}

9. ``Analytic solution for grand confluent hypergeometric function'' \cite{Chou2012i}-apply the three term recurrence formula, and formulate the exact analytic solution of grand confluent hypergeometric function (including all higher terms of $A_n$'s). Replacing $\mu $ and $\varepsilon \omega $ by 1 and $-q$ transforms the grand confluent hypergeometric function into the Biconfluent Heun function.
\vspace{3mm}

10. ``The integral formalism and the generating function of grand confluent hypergeometric function'' \cite{Chou2012j}--apply the three term recurrence formula, and construct an integral formalism and a generating function of grand confluent hypergeometric function (including all higher terms of $A_n$'s). 
\section*{Acknowledgment}
I thank Bogdan Nicolescu. The discussions I had with him on number theory was of great joy.  
\vspace{3mm}

\bibliographystyle{model1a-num-names}
\bibliography{<your-bib-database>}
 
\end{document}